\documentclass[letterpaper]{article} 
\usepackage[submission]{aaai24}  
\usepackage{times}  
\usepackage{helvet}  
\usepackage{courier}  
\usepackage[hyphens]{url}  
\usepackage{graphicx} 
\usepackage{natbib}  
\usepackage{caption} 
\usepackage{algorithm}
\usepackage{algorithmic}
\usepackage{ltl}
\usepackage{xspace}
\usepackage{float}
\usepackage{newfloat}
\usepackage{listings}
\DeclareCaptionStyle{ruled}{labelfont=normalfont,labelsep=colon,strut=off} 
\floatstyle{ruled}
\newfloat{listing}{tb}{lst}{}
\floatname{listing}{Listing}
\usepackage{tcolorbox} 
\usepackage{listings} 
\usepackage{xcolor}    
\usepackage{subcaption}
\usepackage{amsmath}

\usepackage{booktabs}

\newcommand{\upd}[2]{\ensuremath{[ \hspace{0.5pt} #1 \leftarrowtail \, #2 \hspace{0.5pt} ]}}
\newcommand{\name}[1]{{\text{\texttt{#1}}}}

\newcommand{\term}{\ensuremath{\tau}}
\newcommand{\fterm}{\ensuremath{\term_{F}}\xspace}
\newcommand{\pterm}{\ensuremath{\term_{P}}\xspace}

\newcommand{\terms}{\ensuremath{\mathcal{T}}\xspace}
\newcommand{\pterms}{\ensuremath{\terms_{\!P\hspace{-0.5pt}}}\xspace}
\newcommand{\fterms}{\ensuremath{\terms_{\!F\hspace{-0.5pt}}}\xspace}

\newcommand{\sep}{\ensuremath{\quad | \quad}}

\newcommand{\sats}{\ensuremath{\;\vDash_{\!\langle \hspace{-1pt} \cdot \hspace{-1pt} \rangle}}}

\newcommand{\branch}[2]{\ensuremath{#1 \! \wr \hspace{-0.6pt} #2}}
\newcommand{\assign}[1]{\ensuremath{\langle #1 \rangle}}
\newcommand{\functions}{\ensuremath{\mathcal{F}}\xspace}
\newcommand{\fnames}{\ensuremath{\mathbb{F}}\xspace}

\title{Combining LLM Code Generation with Formal Specifications and Reactive Program Synthesis}
\author{
    Written by AAAI Press Staff\textsuperscript{\rm 1}\thanks{With help from the AAAI Publications Committee.}\\
    AAAI Style Contributions by Pater Patel Schneider,
    Sunil Issar,\\
    J. Scott Penberthy,
    George Ferguson,
    Hans Guesgen,
    Francisco Cruz\equalcontrib,
    Marc Pujol-Gonzalez\equalcontrib
}
\affiliations{
    \textsuperscript{\rm 1}Association for the Advancement of Artificial Intelligence\\

    1900 Embarcadero Road, Suite 101\\
    Palo Alto, California 94303-3310 USA\\
    proceedings-questions@aaai.org
}

\begin{document}

\maketitle

\begin{abstract}
  In the past few years, Large Language Models (LLMs) have exploded in usefulness and popularity for code generation tasks.
  However, LLMs still struggle with accuracy and are unsuitable for high-risk applications without additional oversight and verification.
  In particular, they perform poorly at generating code for highly complex systems, especially with unusual or out-of-sample logic.
  For such systems, verifying the code generated by the LLM may take longer than writing it by hand.
  We introduce a solution that divides the code generation into two parts; one to be handled by an LLM and one to be handled by formal methods-based program synthesis.
  We develop a benchmark to test our solution and show that our method allows the pipeline to solve problems previously intractable for LLM code generation.
\end{abstract}


\section{Introduction}
The potential for LLMs to increase the productivity of software engineers through code-generation has been well-demonstrated through popular benchmarks such as HumanEval \cite{humaneval} or SWE-Bench \cite{swebench}
However, despite their strong code-writing capabilities, the adoption of LLMs in writing code for mission-critical systems hinges on the inability of LLMs to write code with formal correctness guarantees.
In particular, using LLMs for the generation of code bases containing potentially hundreds of thousands of lines requires that the generated code then be manually verified, a time-consuming task that negates many of the advantages of LLM code generation.


In this work, we propose that LLM code generation can be combined with the large body of existing work in program synthesis in formal methods~\cite{alur2013syntax, gulwani2011automating,jacobs20185th}, both to increase correctness of generations as well as to decrease the lines of code that must be manually verified.
At a high level, we use LLMs to generate formal specifications from natural language, we then use program synthesis to generate code that is ``correct-by-construction''.
A major challenge to this approach is that program synthesis from formal specifications generally targets a small, fixed grammar, meaning we are not able to leverage the flexibility of LLM code generation.
Our key insight is to use formal specification languages that allow for ``holes'' in the generated code that can be later filled in with LLM code generation.
In this way, we use program synthesis from formal synthesis only for the generation of the structure of the code base, then we can fill in the details with an LLM in such a way that the structural guarantees are still valid regardless of the output of the LLM. 

Specifically, we use Temporal Stream Logic, which allows authors to specify brief logical constraints on a system's behavior to generate reactive systems with complexity surpassing what a maintainer could easily implement, and augmenting it with the flexibility and reasoning capabilities of LLMs.
In this way, we open the door to the creation of systems that can dynamically generate trustworthy code for high-risk reactive systems, tackling the issue of trustworthiness of LLMs and the drawback of complex specs of TSL. 

In summary, we identify the key contributions of this work as follows:

\begin{itemize}
    \item Propose a framework for combining formal specification-based program synthesis with LLM code generation to reduce the amount of generated code that must be verified.
    \item An instantiation of this framework into a code generation pipeline using Temporal Stream Logic.
    \item An evaluation of our system on a two set of reactive program synthesis benchmarks.
\end{itemize}


\begin{figure*}[!t]
  \centering
  \includegraphics[scale=0.95]{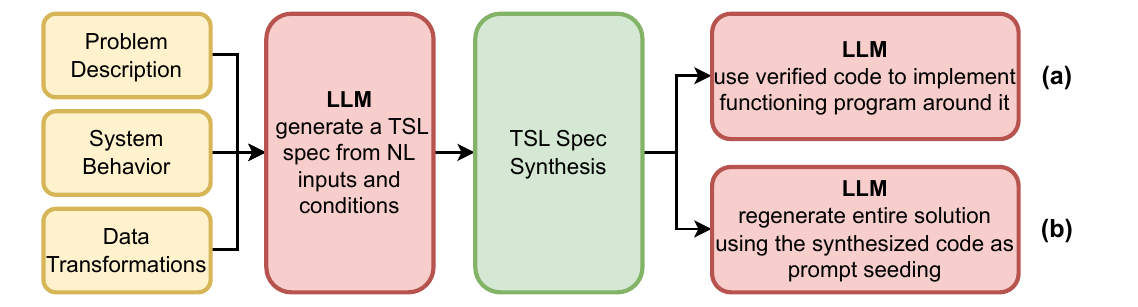}
  \caption{Overview of the full TSL + LLM code generation pipeline. The pipeline turns natural language into executable code. Pipeline \textbf{(a)} implements wrapper code to interact with the synthesized code similar to an API. Pipline \textbf{(b)} uses the synthesized code to seed an LLM prompt to generate the program described by the NL descriptions.}
  \label{fig:pipeline}
\end{figure*}

\section{Related Work} 
\subsubsection{LLM Code Generation}
LLM code generation has advanced significantly to where models are able to outperform humans at competitive programming benchmarks such as HumanEval~\cite{humaneval}. 
Popular models like GPT-4~\cite{gpt4} and open source models like CodeLlama~\cite{codellama} have made code generation accessible to researchers and industry to tackle problems that were previously out of scope for traditional program synthesis techniques~\cite{swebench}. 
In Magicoder, \cite{wei2024magicoder} find that seeding code generation with randomly sourced code snippets improves the quality of generated code significantly, indicating that using synthesized code to seed generation may be a promising strategy.
However, most benchmarks used for code evaluation~\cite{humaneval,wei2024magicoder,codellama} focus on short competitive coding problems, which require relatively few lines of code and are relatively easily verifiable. 
Benchmarks such as SWE-Bench~\cite{swebench} are larger in scope and are markedly harder for LLMs to solve correctly.
In the case that the generated code requires manual verification beyond the provided tests, the use of code generation becomes an exhaustively time-consuming process. 

\subsubsection{Formal Synthesis}
The field of Reactive Synthesis has seen tremendous progress since its first formalization as the Church synthesis problem~\cite{church1962logic}.
Most commonly specifications are provided through Linear Temporal Logic (LTL), and the goal is to synthesize a system that reacts to an infinite stream of inputs.
LTL aims to place logical, mathematically rigorous guarantees on a system's behavior, to verify an existing system or generate a new one. 
A primary example of the successful use of LTL for reactive synthesis is the AMBA bus protocol~\cite{bloem2014parameterized}. 
Since then, the field has seen significant milestones in education~\cite{ma2023using}, FPGA game development~\cite{geier2019syntroids}, music~\cite{choi2021program}, and interactive animations~\cite{rothkopf2023towards}. 
Other formal synthesis systems like Temporal Stream Logic (TSL) \cite{tsl} provide important advancements, separating data and control, and utilizing function predicate terms which simplify code synthesis but make the final implementation harder for users, who are required to implement the function and predicate terms as well as handle integration of the synthesized code into larger projects \cite{tsl}.

TSL introduces predicate terms, $ \pterm \in \pterms $, which are used to make observations on the environment, and function terms $ \fterm \in \fterms $, which are used to construct output values.
These predicate terms enable users to decouple the data and control aspects of a system, encapsulating functionality within functions and predicates that are not pertinent to the specification. 
As a result, the specification only needs to address the essential parts of the system required to ensure the desired guarantees.
In this work, the separation of data and control allows us to leverage the LLM's flexibility for code generation of function and predicate terms, while also allowing Reactive Synthesis with TSL to handle the logical reasoning task of temporal structure that is less well-suited to an LLM. 


\subsubsection{Using LLMs and Formal Synthesis} 
In pursuit of enhancing the capabilities of formal and LLM systems, researchers have explored combining both approaches. The works of nl2spec \cite{cosler2023nl2spec} and Lang2LTL \cite{liu2022lang2ltl}, have explored the transformation of natural language descriptions into LTL specifications. 
These approaches aim to bridge the gap between informal user requirements and formal temporal logic specifications, enabling a more accessible and intuitive method for defining system behaviors and properties. \cite{rothkopf2024enforcing} explored how reactive synthesis can be leveraged to enforce temporal constraints on content generated by LLMs, but do not explore LLM code generation. 

\subsubsection{Zero, few-shot learning and in context learning (ICL)} are exciting new abilities of powerful LLMs, that enable performance similar to fine-tuned models without costly dataset creation and fine-tuning processes \cite{brown2020languagemodelsfewshotlearners}. Zero and few-shot learning can boost model performance significantly; \cite{brown2020languagemodelsfewshotlearners} find that models demonstrate an ability to learn about a problem from structured prompts, we believe that the highly structured nature of TSL \cite{tsl} will help LLMs correctly convert NL to TSL. Additionally, we leverage the ability of models to do ICL and learn from the structure of inputs to boost performance \cite{min2022rethinkingroledemonstrationsmakes}. 




\begin{figure*}[t] 
    \centering
    \begin{subfigure}[t]{0.48\textwidth}
        \centering
        \begin{tcolorbox}[colback=yellow!12!white, colframe=yellow!50!black, title=Structured NL Prompt Sample, width=\textwidth]
        \small
        \begin{lstlisting}[breaklines=true, basicstyle=\footnotesize\ttfamily]
You can assume that eventually every truck will make a request. Further, guarantee that:


1. for each truck 1..3, if the truck makes a request, then eventually it will be given a grant.
2. If the coinflip between truck 2 and 3 resolves to true and truck 1 is granted the road, then truck 2 will not be granted the road until truck 3 is granted.
3. If truck 1 is given a grant, then truck 2 won't be until truck 3 is.
        \end{lstlisting}
        \end{tcolorbox}
        \caption{Example of a structured NL prompt that is used to prompt the LLM to generate a TSL specification. This example does not include the few-shot preamble used in prompting the LLM to generate (b).}
        \label{fig:strc_nl}
    \end{subfigure}
    \hfill
    \begin{subfigure}[t]{0.48\textwidth}
        \centering
        \begin{tcolorbox}[colback=orange!12!white, colframe=orange!50!black, title=TSL specification Sample, width=\textwidth]
        \small
        \begin{lstlisting}[breaklines=true, language=Python, basicstyle=\footnotesize\ttfamily]
# Assumptions: describes inputs from the environment
always assume {
  F (r t);
}

# Guarantees: describes how the agent react to those inputs
always guarantee {
    (r 1) -> F ([ g <- 1 ]);
    (r 2) -> F ([ g <- 2 ]);
    (r 3) -> F ([ g <- 3 ]);
    ((p 2 3) && [ g <- 1 ] ->  ! ([ g <- 2 ]) W [ g <- 3 ]);
    ([ g <- 1 ] -> ((! ([ g <- 2 ]) W [ g <- 3 ])));
}

        \end{lstlisting}
        \end{tcolorbox}
        \caption{Example of a TSL specification that describes an arbiter that can handle three requesters. This specification is semantically identical to the NL description in (a).}
        \label{fig:ex1}
    \end{subfigure}
    \caption{Examples of the TSL specification interface in both natural language and formal logic.}

\end{figure*}

\section{System Overview}\label{sec:system}
Our system is a code generation pipeline as shown in Fig.~\ref{fig:pipeline}, which leverages LLMs to generate TSL specs, and then utilizes correct-by-construction synthesized code to complete critical aspects of data flow control problems such as arbiters or other reactive systems. 
Our system uses LLMs to overcome the burden of writing TSL specifications and leverages the correct-by-construction nature of synthesized code to reduce the amount of unverified code in mission-critical applications.
Moreover, by abstracting complex state behaviors away from the LLM, we can use our pipeline to generate controllers for highly complex multi-agent environments. 

Our pipeline is composed of the following steps: 
\begin{enumerate}
\item Inputs: 
    \begin{enumerate}
        \item A high-level, natural language summary of the problem (Problem Description in Fig.~\ref{fig:pipeline}).
        \item A more detailed, natural language description of the most important assumptions and guarantees needed in the spec (System Behavior in Fig.~\ref{fig:pipeline}).
        \item A separation of data and control in the form of function and predicate terms. The specification should use these to encapsulate logic not relevant to the assumptions and guarantees (Data Transformations in Fig.~\ref{fig:pipeline}).
    \end{enumerate}
\item Using prompt engineering, we utilize an LLM to convert the NL inputs into a TSL specification
\item We use formal methods-based program synthesis to solve the TSL specification and synthesize correct-by-construction code.
\item Using an LLM, we integrate the formally verified code into the larger codebase or project. We capitalize on the unique nature of TSL that separates data and control to generate lower-risk encapsulating logic using an LLM to integrate the verified code, thereby reducing the burden of verification on human developers.
\item Alternatively, we adopt the prompt seeding technique of Magicoder \cite{wei2024magicoder} and use the synthesized code to request the LLM to generate from the ground up a working solution. When implementing real world complex decision systems, this approach will seed the LLM prompt with relevant code for the control logic that is guaranteed to be correct, potentially boosting the rate of generating correct code for the whole system.
\end{enumerate}

\subsection{Natural Language (NL) to TSL}
An important part of our pipeline revolves around an NL to TSL conversion handled by the LLM. We leverage ICL and few-shot prompting to create TSL prompts from detailed NL inputs. We use a prompting template that contains several NL to TSL examples, as well as explicit explanations about TSL terms and statements. We find that GPT-4 \cite{gpt4} can perform very well using few shot prompting. This is likely due to the highly structured and logical nature of TSL, with which models can perform ICL exceptionally well~\cite{min2022rethinkingroledemonstrationsmakes,brown2020languagemodelsfewshotlearners}. 

\begin{figure}[!h]
  \centering
  \includegraphics[scale=0.7]{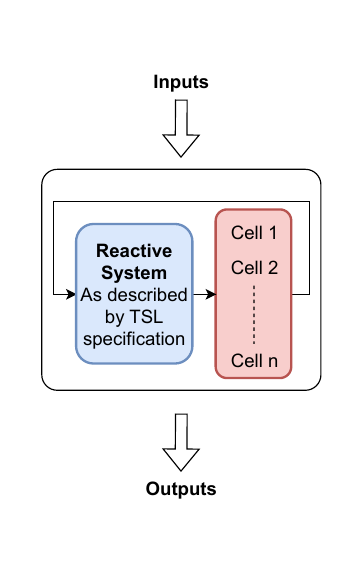}
  \caption{Overview of a reactive system as implemented in a formal TSL specification.}
  \label{fig:reactive_sys}
\end{figure}

TSL's structure allows for a exceptionally clean logical separation of data and control in a program. As shown in Fig. \ref{fig:reactive_sys}, IO streams are separate from the reactive system, as seen in Fig. \ref{fig:strc_nl} and Fig. \ref{fig:ex1}, we define function and predicate terms, $r,\ g$ to handle state changes and data flow in the system, giving greater flexibility to the LLM in creating TSL specifications as well as facilitating integration into full programs. 
The flexibility that function and predicate terms give TSL specifications \cite{tsl}, is likely a factor that facilitates ICL in our tasks, since the LLM can leverage its coding and reasoning strengths to a greater extent in the NL to TSL step~\cite{min2022rethinkingroledemonstrationsmakes,brown2020languagemodelsfewshotlearners}. 
The utility of the function and predicate terms then carries over to the final code implementation step shown in Fig.~\ref{fig:pipeline}, where the LLM can leverage the flexibility of the data control separation in the synthesized code to easily generate wrapper code for the synthesized reactive system. 
Due to the pure functional (side-effect free) nature of the wrapper code, function and predicate terms resolve to functions with clearly defined functionality; this structure can help to make verification a more tractable task.

\begin{figure*}[!t]
  \centering
  \includegraphics[scale=0.75]{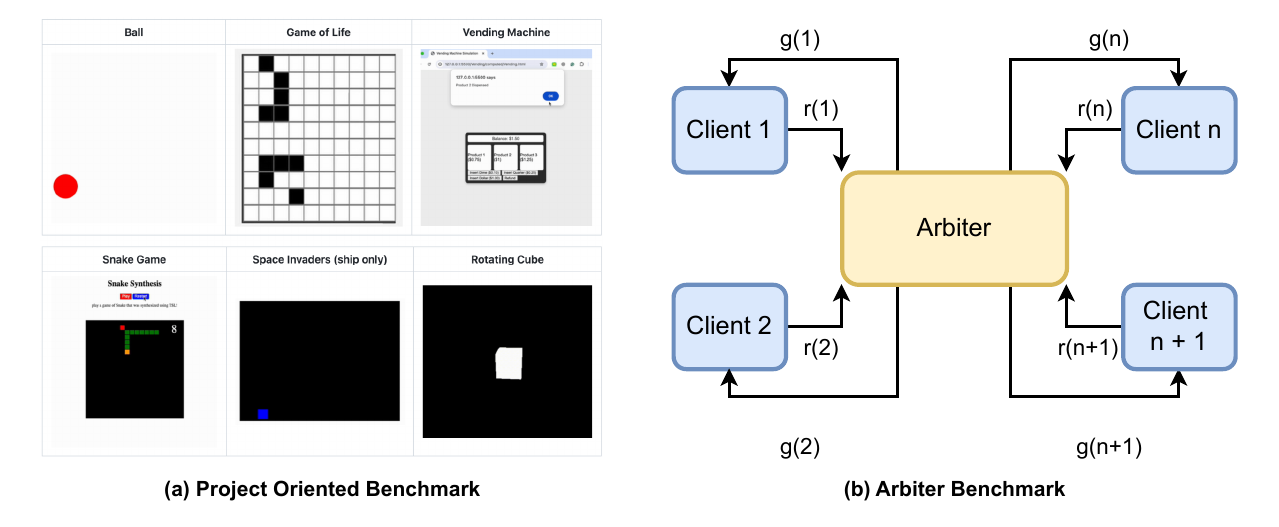}
  \caption{Six different project-oriented coding benchmarks that require working implementation of reactive applications and games.}
  \label{fig:benchmarks}
\end{figure*}

\subsection{Formal Verification Assisted Code Generation}
Our pipeline aims to reduces the number of lines of unverified code in projects by removing the burden of generation from the LLM and moving it to the formal synthesis procedure.
We do this through the introduction of function and predicate terms that manipulate data at a single point in time.
These predicate terms enable users to decouple the data and control (temporal) aspects of a system, encapsulating data transformations within functions and predicates that are not pertinent to the temporal specification~\cite{tsl}. 
As a result, the scope of the specification is reduced to only address the temporal aspect of the system, as opposed to, for example LTL, where both data and control are the captured by the specification and are the responsibility of the synthesis engine.

This is achievable due to TSL's introduction of specification-level predicate and function terms. 
Predicate terms, $\pterm \in \pterms$, are used to make observations on the environment, and function terms, $\fterm \in \fterms$, are used to construct output values, where $ \name{s}_{\name{i}} \in \mathbb{I} \cup \mathbb{C}$ is an input stream or cell value, and $ \name{s}_{\name{o}} \in \mathbb{O} \cup \mathbb{C} $ is an output stream or cell value. 
Together, all the available predicate names $\name{p}$ and all the available function names $\name{f}$ form the set of function symbols $\fnames$. 
A TSL formula describes a system that consumes input $\mathcal{I} = \mathbb{I} \cup \mathbb{C}$ and produces output $\mathcal{O} = \mathbb{O} \cup \mathbb{C}$ as shown in Eq. (\ref{eq:1}). As shown in Fig. \ref{fig:reactive_sys} the reactive system converts an input stream into an output stream, using cells to track states.

\begin{equation}
\label{eq:1}
  \begin{array}{rl}
  \pterm \;:= & \name{p}~\;\fterm^{0}~\;\fterm^{1}~\;\ldots~\;\fterm^{n-1} \\[0.3em]
  \fterm  \;:= & \name{s}_{\name{i}} \sep \name{f}~\;\fterm^{0}~\;\fterm^{1}~\;\cdots~\;\fterm^{n-1} \\[0.3em]
  \varphi  \; := & \pterm \ \, | \ \, \upd{\name{s}_{\name{o}}}{\fterm} \ \, | \ \, \neg \varphi \ \, | \ \, \varphi \wedge \varphi \ \, | \ \, \LTLnext \varphi \ \, | \ \, \varphi \LTLuntil \varphi \\ \\
  \end{array}
\end{equation}

The realizability problem of TSL is stated as follows: given a TSL formula~$ \varphi $, is there a strategy~$ \sigma \in \mathcal{I}^{+} \to \mathcal{O} $ mapping a finite input stream (since the beginning of time) to an output (at each particular timestep),
such that for any infinite input stream $ \iota \in \mathcal{I}^{\omega} $,
and every possible interpretation of the function symbols (where an interpretation is some concrete implementation in code) $ \assign{\cdot}: \fnames \to \functions $, 
the execution of that strategy over the input $ \branch{\sigma}{\iota} $ satisfies $ \varphi $, i.e.,

\begin{equation}
  \exists \sigma \in \mathcal{I}^{+} \to \mathcal{O}. \ \, 
  \forall \iota \in \mathcal{I}^{\hspace{0.2pt}\omega}. \ \, 
  \forall \assign{\cdot} : \fnames \to \functions. \ \, 
  \branch{\sigma}{\iota}, \iota \sats \varphi
\end{equation}

If such a strategy~$ \sigma $ exists, we say that $ \sigma $ realizes $ \varphi $.
The key insight here is that in TSL we are universally quantifiying over implementations of predicate and function terms.
The specification $\varphi$ only describes a temporal relation of predicate evaluations to function applications--abstracting away from what these predicates and functions do to any underlying data.
In TSL synthesis, this model $\sigma$ can be turned into a block of program code that describes a Mealy machine~\cite{mealy1955method}, where the transitions represent function and predicate terms. 
These assurances allow us to treat synthesized code as verified in large codebases. The user only needs to manually check the correctness of the specification is correct.

\subsection{Simplifying Long Context Problems}

Using TSL specifications allows us to shorten the context of NL instructions describing reactive systems. 

TSL's data and control separation through its universal quantification of function and predicate terms is particularly well-suited for long-context LLM generation. TSL lets us encode complex conditional statements in a compact syntactical representation.
In this way, the LLM generated TSL specifications are significantly easier to verify than long LLM-generated code bases, increasing efficiency and trust in the code. 

This separation enables developers to first focus on when their system should execute certain behaviors, and leave the question of how those behaviors should be implemented for a later step in the development process.
Traditionally, the control is synthesized from TSL and the end-user implements the data transformations manually.
We propose automating both processes using LLMs.


\section{Evaluation}
We evaluate our approach in two different contexts. First, we seek to quantify the improvements of our TSL-enhanced generation over traditional LLM-only code generation. We measure and compare the number of unverified lines of code written by either method over a series of project-driven benchmarks whose goal is to implement simple reactive programs as demonstrated in \ref{fig:benchmarks}(a). Next, we compare the benefits of using our TSL pipeline in generating solutions to complex arbiter problems as shown in \ref{fig:benchmarks}(b); we seek to progressively explore the limits of LLM-only generated arbiters as opposed to TSL pipelines. We depart from traditional code generation testing frameworks~\cite{humaneval} and design our tasks to be very hard for LLMs to solve, following \cite{swebench} in SWE-Bench. 

\subsection{Testing Methodology}

\subsection{Task Oriented Code Generation}
The task-oriented code generation metric seeks to quantify the reduction in lines of unverified code present in a codebase when using TSL pipelines. We count the lines of code written for a pass at $k$ condition to create a working implementation of the program for both the TSL pipeline and the LLM-only approach. To ensure consistency across both approaches, we use the same structured NL prompt shown in Fig. \ref{fig:pipeline} in both runs, prompting the LLM to generate a working implementation from scratch in one case, and a TSL spec in the other. In our experiments, we include runs where the TSL specification fails to compile. We do this to explore the effects of changing the LLM architecture on few-shot prompting with examples. In real-world applications, rejecting unsynthesizable TSL specifications and using techniques like reprompting can lead to much better success rates. As shown in Fig \ref{fig:pipeline}(b), we take inspiration from \cite{wei2024magicoder} who show that seeding LLM code generation with random code significantly improves output. We expect that using verified and task-specific code to further improve the outputs of the pipeline over LLM-only generation.

We implement this task with a full generation objective, that tracks the total lines of code needed to generate a working implementation of the task. As such we are able to demonstrate differences in the lines of unverified code introduced into codebases as well as examine the success rate of various methods when generating complex programs with many lines of code.

\subsection{Complex Arbiter Problem Solution Generation}
We evaluate the improvements of our approach on a scaling arbiter benchmark as shown in \ref{fig:benchmarks}(b). This task separates data and control in a resource management game, in which the generated code must track resource requests and requesters and follow these rules:

\begin{itemize}
    \item Every request of a client is eventually granted by the arbiter.
    \item The arbiter never two grants at the same time.
    \item The arbiter only grants a client $i$ if it has an open request.
    \item A client may only pose a request if it has no open request.
\end{itemize}

This task is designed to leverage the verified nature of the synthesized lines. Since the number of lines required to design an arbiter scales with the number of conditions in the arbiter, we seek to examine the effectiveness of TSL-enhanced pipelines in generating working arbiters, as well as the ability of TSL to cut down on the number of unverified lines introduced into the code base.

\begin{table*}[!t]
  \caption{Task Oriented Code Generation Performance for TSL pipeline, LLM only code generation, and prompt seeding with synthesized code. We compare success rate (sr), average lines of code, and the percentage of unverified lines of code across these tasks.}
  \label{tab:task_performance}
  \vspace{-0.1in}
  \centering
  \begin{tabular}{lcccccc}
    \toprule
    Tasks & sr $5$ & sr $10$ & sr $15$ & Avg. Total Lines & Avg. $\#$ Unverified Lines \\
    \midrule
    Ball\_TSL & $20\%$ & $10\%$ & $13\%$ & $249$ & 78  \\
    Game of Life\_TSL & $60\%$ & $50\%$ & $73\%$ & 124 & 98  \\
    Vending Machine\_TSL & $100\%$ & $100\%$ &$\textbf{100\%}$& $192$ & 71  \\
    Space Invaders\_TSL & $0\%$ & $0\%$ & $6\%$ & $88$ & 68  \\
    Rotating Cube\_TSL & $40\%$ & $40\%$ & $33\%$ & $108$ & 75 \\
    \midrule
    Ball\_LLM & $20\%$ & $10\%$ & $13\%$ & $55$ & - \\
    Game of Life\_LLM & $40\%$ & $50\%$ & $40\%$ & $87$ & - \\
    Vending Machine\_LLM & $100\%$ & $100\%$ &$\textbf{100\%}$& $70$ & - \\
    Space Invaders\_LLM & $40\%$ & $30\%$ &$\textbf{27\%}$ & $68$ & - \\
    Rotating Cube\_LLM  & $80\%$ & $60\%$ &$\textbf{80\%}$& $92$ & - \\
    \midrule
    Ball\_Regen & $100\%$ & $100\%$ & $\textbf{100\%}$& $49$ & - \\
    Game of Life\_Regen & $60\%$ & $70\%$ & $\textbf{80\%}$& $98$ & - \\
    Vending Machine\_Regen & $100\%$ & $100\%$ &$\textbf{100\%}$&$63$& - \\
    Space Invaders\_Regen & $0\%$ & $0\%$ & $6\%$ & $68$ & - \\
    Rotating Cube\_Regen  & $80\%$ & $60\%$ & $\textbf{80\%}$ & $45$ & - \\
    \bottomrule
  \end{tabular}
\end{table*}


\section{Results}

\subsection{Trusted Code and Prompt Seeding}
The first task measures the reduction in lines of unverified code as well as the success rate of full generation both via pipeline and LLM-only methods. Our results indicate that the two ways of using the TSL pipeline both have advantages and disadvantages. On the one hand, using the TSL pipeline without prompt seeding as shown in Fig. \ref{fig:pipeline}(a) results in a large portion of verified lines implementing the system which do not need to be verified. As seen in Table \ref{tab:task_performance} the given benchmark tasks are too simple to leverage this key benefit of the pipeline. We therefore demonstrate that the pipeline shown in \ref{fig:pipeline}(b) can beat a GPT-4 \cite{gpt4} baseline at generating complex programs. On challenging tasks like implementing a working implementation of Conway's game of life, the prompt seeding approach achieves a $0.80$ success rate as opposed to GPT-4's $0.40$ success rate at $15$ tries. 

The rather ambiguous results of the pipeline shown in Fig \ref{fig:pipeline}(a) suggests that even longer and more complex problems are needed to leverage the full potential of formally verified code in codebases, which we explore further in our other experiments. However, we are able to demonstrate strong improvements at complex tasks when using the prompt seeding approach of the pipeline shown in Fig \ref{fig:pipeline}(b). Nonetheless, Table \ref{tab:task_performance} shows that pipeline (a) can match GPT-4 only generation in some tasks, even beat it by $0.33$ at Conway's game of life. Our results therefore indicate that using LLMs to generate code deployed in high-risk decision making environments, requiring hundreds or possibly thousands of guarantees and logical conditions, may very well stand to benefit from pipelines that incorporate TSL.

Our method does introduce certain drawbacks, especially through the TSL specification. In certain cases, the LLM was unable to produce a synthesizable TSL specification which led to poor performance compared to LLM only code generation. This weak performance demonstrates the limits of few shot prompting when requiring the LLM to generate a syntactically complicated and unknown language. 

\begin{table}[!h]
  \caption{Arbiter Problem Model Performance Comparison on Success Rate (sr). We run each trial $25$ times and show the number of verified synthesized lines as well as the number of generated lines to use the synthesized system.}
  \label{tab:arbiter}
  \centering
  \begin{tabular}{lccc}
    \toprule
    \textbf{Metric} & \multicolumn{3}{c}{\textbf{Condition Count}} \\
    \cmidrule(lr){2-4}
     & \textbf{10} & \textbf{20} & \textbf{30} \\
    \midrule
    TSL pipeline (sr)   & $61.5\%$ & $45.4\%$ & $16.7\%$ \\
    LLM pipeline (sr)   & $0\%$ & $0\%$ & $0\%$ \\
    Average line count (TSL) & 22 & 31 & 40 \\
    \bottomrule
  \end{tabular}
\end{table}

\subsection{Solving Long Context Problems}
Our results for this task clearly show that a state-of-the-art LLM like GPT-4 is unable to solve complex decision problems as described in Fig \ref{fig:strc_nl} and \ref{fig:ex1}, whilst our new TSL enhanced pipeline shown in Fig \ref{fig:pipeline}(a) is able to enhance the performance of the LLM. Crucially, the success rate of GPT-4 at these arbiter problems is $0$, whilst the TSL pipeline is able to solve between $62\%$ and $17\%$ of arbiter problems with $10,\ 20$ or $30$ conditions. TSL allows the LLM to solve the problem, and this solution is practical because the number of generated lines does not grow to the point that each cannot be verified by hand. As shown in Table \ref{tab:arbiter}, the number of generated lines grows linearly with the number of conditions; the interacting requirements do not create exponential complexity that would cause it to be impractical to review. 

These results demonstrate how using TSL-enhanced generation can introduce the advantages of LLM code generation to a whole new field of complex decision systems. 

\section{Conclusion}
In this paper, we demonstrate the possibility of building a code generation pipeline by moving key system logic to formal methods-based program synthesis without sacrificing the strong code-generating capabilities of state-of-the-art LLMs. 
Additionally, we show that, when not concerned with verification, LLM code generation can be significantly improved when using TSL-based prompt seeding. 

In conclusion, our work has demonstrated that adding formal verification methods can significantly enhance and improve traditional code generation pipelines. Our results show that across various tasks, we can reduce the lines of unverified code in applications, making this pipeline more desirable for high-risk applications. Moreover, temporal logic allows this pipeline to generate controllers for highly complex arbitration problems. Our pipeline relies on ICL and few-shot prompting to convert NL to TSL, opening interesting future work into fully integrated and fine-tuned pipelines.


\bibliography{aaai24}
\end{document}